\documentclass[conference]{IEEEtran}
\IEEEoverridecommandlockouts

\usepackage{cite}

\ifCLASSINFOpdf
\else
\fi
\usepackage[cmex10]{amsmath}

\usepackage[T1]{fontenc}
\usepackage{ae}
\usepackage{aecompl}
\usepackage{amssymb}
\newtheorem{definition}{Definition}[section]
\newtheorem{theorem}{Theorem}[section]
\newtheorem{lemma}{Lemma}[section]
\newtheorem{remark}{Remark}[section]

\newtheorem{corollary}{Corollary}[section]

\begin{document}
%
\title{Information Spectrum Approach to Overflow Probability of Variable-Length Codes with Conditional Cost Function}
\author{\IEEEauthorblockN{Ryo NOMURA}
\IEEEauthorblockA{School of Network and Information\\
Senshu University, Japan.\\
E-mail: nomu@isc.senshu-u.ac.jp}
\and
\IEEEauthorblockN{Toshiyasu MATSUSHIMA}
\IEEEauthorblockA{School of Fundamental Science and Engineering\\ Waseda University, 
Japan.}
\thanks{This research was supported in part by JSPS Grant-in-Aid for Young Scientists (B) No. 23760346.}
}


%


\maketitle

\begin{abstract}
Lossless variable-length source coding with unequal cost function is considered for {\it general} sources.
In this problem, the codeword cost instead of codeword length is important. The infimum of average codeword cost has already been determined for {\it general} sources.
We consider the overflow probability of codeword cost and determine the infimum of achievable overflow threshold.
Our analysis is on the basis of {\it information-spectrum} methods and hence valid through the {\it general} source.
\end{abstract}


%
\IEEEpeerreviewmaketitle

\section{Introduction}
Lossless variable-length coding problem is quite important not only from the theoretical viewpoint but also from the viewpoint of its practical applications.
The most fundamental criterion to evaluate the performance of variable-length codes is the average codeword length. 
The overflow probability of codeword length is one of other criteria, which denotes the probability of codeword length per symbol being above a threshold $R >0$.
The overflow probability has been investigated in several contexts.
Merhav  \cite{Merhav91:overflow} has determined the optimal exponent of the overflow probability given $R$ for unifilar sources.

On the other hand, as is well known, if we impose {\it unequal} costs on code symbols, 
we have to consider the codeword {\it cost}, instead of codeword length.
The average codeword {\it cost}, which is a generalization of the average codeword length, have been first analyzed by Shannon \cite{Shannon}. 
Karp \cite{Karp} has given the variable-length code, which minimizes the average codeword cost. 
The infimum of the average codeword cost has been determined by Krause \cite{Krause} for i.i.d. sources and extended to {\it general} sources by Han and Uchida \cite{Han00:cost}.
Among others, Uchida and Han \cite{Uchida01:overflow} have proposed the overflow probability of codeword {\it cost}. They have considered the overflow probability as the probability of codeword {\it cost} per symbol being above a threshold. Then, they have shown the infimum of achievable threshold, where achievable threshold means that there exists a code whose overflow probability of codeword cost decreases with given exponent $r$.

In this paper, we also deal with the overflow probability of codeword {\it cost}.
In particular, we consider the  $\varepsilon$-achievable threshold, which means that there exists a variable-length code, whose overflow probability is smaller than or equal to $\varepsilon$.
Then, we first determine the infimum of $\varepsilon$-achievable threshold for {\it general} sources.
Next, we consider the {\it second-order} achievable threshold for {\it general} sources.
The finer evaluation of the achievable rates, called the second-order achievable rates, have been analyzed in several contexts\cite{Kon,Hayashi,Polyanskiy2010,NH2011}. 
Analogously to these settings, we define the {\it second-order} achievable threshold on the overflow probability and derive the infimum of the {\it second-order} achievable threshold.
Nomura and Matsushima \cite{Nomura_SITA2011} have first derived this quantity in the case that the cost function is assumed to be memoryless.
In this paper, we treat the conditional cost function to generalize this result.
As a result, the similar analysis with \cite{Nomura_SITA2011}, which is on the basis of the {\it information-spectrum} methods,  is also efficient in this case.
These results show that the {\it information-spectrum} methods are still effective in analyses of the overflow probability in variable-length coding problem.
%
%
%
%
%
\section{Preliminaries}
\subsection{Variable-length codes with cost for general source}
The {\it general} source is defined as an infinite sequence
\[
{\bf X} = \left\{ X^n = \left( X_1^{(n)}X_2^{(n)}\cdots X_n^{(n)}  \right) \right\}_{n=1}^\infty
\]
of $n$-dimensional random variables $X^n$, where each component random variable $X_i^{(n)}$ takes values in a {\it countable} set ${\cal X}$.

The variable-length codes are characterized as follows.
Let
\[
\varphi_n : {\cal X}^n \rightarrow {\cal U}^{\ast}, \ \  \psi_n : \{\varphi_n({\bf x})\}_{{\bf x} \in {\cal X}^n} \rightarrow {\cal X}^n,
\]
be a variable-length encoder and a decoder, respectively, where ${\cal U} = \{ 1,2,\cdots,K\}$ 
is called the code alphabet and ${\cal U}^{\ast}$ is the set of all finite-length strings over ${\cal U}$ excluding the null string.

We consider the situation that there are {\it unequal costs} on code symbols. Let us define the cost function $c: {\cal U}^{\ast} \to (0, +\infty)$ considered in this paper.
The cost $c(u^l)$ of a sequence $u^l \in {\cal U}^l$ is defined by
\[
c(u^l) = \sum_{i=1}^{l}c(u_i|u_1^{i-1}),
\]
where $c(u_i|u_1^{i-1})$ is a conditional cost of $u_i$ given $u_1^{i-1}$ such that 
$0 < c(u_i|u_1^{i-1}) < \infty$ ($\forall i,$$\forall u_i \in {\cal U},$$\forall u_1^{i-1} \in {\cal U}^{i-1}$). 
The conditional {\it cost capacity} $\alpha_c(u_1^{i-1})$ given $u_1^{i-1}$ is defined by the positive unique root $\alpha$ of the equation
\begin{equation} \label{capacity}
\sum_{u_i \in {\cal U}} K^{-\alpha c(u_i|u_{1}^{i-1})} =1.
\end{equation}
In this paper, we assume that the conditional {\it cost capacity} $\alpha_c(u_1^{i-1})$ is independent on $u_1^{i-1}$, more exactly, $\alpha_c(u_1^{i-1}) = \alpha $ holds for all $u_1^{i-1} \in {\cal U}^{i-1}$.

Such a class of cost function has been first considered in \cite{HK97}. Han and Uchida \cite{Han00:cost} also have treated this type of cost function.
We denote 
\[c_{max} = \max_{i,u_i \in {\cal U},u_1^{i-1} \in {\cal U}^{i-1}} c(u_i|u_1^{i-1})\]
 for short.
The variable-length codes considered in this paper, satisfies the prefix condition:
\begin{equation} \label{kraft}
\sum_{{\bf x}\in{\cal X}^n}K^{-\alpha_c c(\varphi_n({\bf x}))} \leq 1.
\end{equation}
Throughout this paper, the logarithm is taken to the base $K$.
%
%
%
%
%
%
%
\subsection{Overflow Probability of Codeword Cost}
The overflow probability of codeword length is defined as follows:
\begin{definition}
Given a threshold $R$, the overflow probability of the variable-length encoder $\varphi_n$ is defined by
\[
\varepsilon_n(\varphi_n, R) = \Pr\left\{ \frac{1}{n}l(\varphi({X^n})) >R \right\},
\]
where $l(\cdot)$ denotes the length of a string.
\end{definition}

In this paper, we generalize the above overflow probability not only to the case for {\it unequal costs }on code symbols but also for finer evaluation of the overflow probability.
To this end, we consider the overflow probability of codeword cost as follows:
\begin{definition}
Given some sequence $\{\eta_n\}_{n=1}^\infty$, where $0 < \eta_n < \infty$ for each $n=1,2,\cdots$, the overflow probability of the variable-length encoder $\varphi_n$ is defined by
\begin{equation}\label{gof}
\varepsilon_n(\varphi_n, \eta_n) = \Pr\left\{ c(\varphi_n({X^n})) > \eta_n \right\}.
\end{equation}
\end{definition}
Since $\{\eta_n\}_{n=1}^\infty$ is an arbitrary sequence, the above definition is general.
In particular, we shall consider the following two types of overflow probability in this paper:
\begin{enumerate}
\item $\eta_n = nR$,
\item $\eta_n = na + \sqrt{n}L.$
\end{enumerate}
\begin{remark}
If we set $\eta_n = nR$ for all $n=1,2,\cdots$, the overflow probability can be written as
\begin{align*}
\varepsilon_n(\varphi_n,nR) &= \Pr\left\{ \frac{1}{n} c(\varphi_n(X^n)) > R \right\}.
\end{align*}
Thus, in the case that $\eta_n = nR$, the overflow probability defined by (\ref{gof}) means the probability that the codeword cost per symbol exceeds some constant $R$.
This is a natural extension of the overflow probability of codeword length analyzed in \cite{Merhav91:overflow} or \cite{Uchida01:overflow} to the overflow probability of {\it codeword cost}.

On the other hand, the finer evaluation of the achievable rates, called the second-order achievable rates, has been investigated in several contexts\cite{Kon,Hayashi,Polyanskiy2010,NH2011}. 
Analogously to their analyses, we evaluate the overflow probability in the {\it second-order} sense.
To do so, we consider the second case: $\eta_n = na+L\sqrt{n}$ for all $n=1,2,\cdots$.
Hereafter, if we consider the overflow probability in the case $\eta_n = na+L\sqrt{n}$, we call it the {\it second-order} overflow probability given $a$ in this paper, while in the first case it is called the {\it first-order} overflow probability.
The {\it second-order} overflow probability given $a$ of variable-length encoder $\varphi_n$ with threshold $L$ is written as
\begin{align*}
\varepsilon_n\left(\varphi_n,na+L\sqrt{n}\right) & = \Pr\left\{\frac{ c(\varphi_n(X^n)) -  n a }{\sqrt{n}}  > L \right\}.
\end{align*}
\end{remark}

In the first case, we are interested in the infimum of threshold $R$ that we can achieve.
\begin{definition} 
Given $0 \leq \varepsilon < 1 $, $R$ is called an $\varepsilon$-achievable threshold for the source if there exists a variable-length encoder $\varphi_n$ such that
\begin{equation*}
\limsup_{n \rightarrow \infty} \varepsilon_n(\varphi_n,nR) \leq \varepsilon.
\end{equation*}
\end{definition}
\begin{definition}
\begin{eqnarray*}
R(\varepsilon|{\bf X})& \stackrel{\mathrm{def}}{=} & \inf \left\{ R| R \mbox{ is an $\varepsilon$-achievable threshold} \right\}.
\end{eqnarray*}
\end{definition}

Also, in the analysis of second-order overflow probability, we define the achievability:
\begin{definition}
Given $0 \leq \varepsilon < 1 $ and $0 < a < \infty$, $L$ is called an $(\varepsilon,a)$-achievable threshold for the source, if there exists a variable-length encoder $\varphi_n$ such that
\begin{equation*}
\limsup_{n \rightarrow \infty} \varepsilon_n\left(\varphi_n,na+L\sqrt{n}\right) \leq \varepsilon.
\end{equation*}
\end{definition}
\begin{definition}
\begin{align*}
L(\varepsilon,a|{\bf X}) \stackrel{\mathrm{def}}{=}  \inf \left\{ L| L \mbox{ is an $(\varepsilon,a)$-achievable threshold} \right\}.
\end{align*}
\end{definition}
%
%
%
%
\section{Key Lemmas}
We show two lemmas that have important roles to derive our main theorems. These lemmas are derived using the {\it information-spectrum} method and hence valid through the {\it general} source.
\begin{lemma} \label{lemma1}
For any general sources ${\bf X}$ and any sequence of positive number $\{\eta_n \}_{n=1}^\infty$, there exists a variable-length encoder $\varphi_n$ that satisfies
\begin{align*}
\varepsilon_n(\varphi_n,\eta_n) < \Pr\left\{ z_n P_{X^n}(X^n)\!\leq\!K^{\!-\alpha_c \eta_n} \right\} + z_n K^{\alpha_c c_{max}+1},
\end{align*}
for $n=1,2,\cdots$, where $\{ z_n \}_{n=1}^\infty$ is a given sequence of an arbitrary number satisfying $z_i > 0$ for $i=1,2,\cdots$ and $\alpha_c$ denotes the {\it cost capacity} defined in (\ref{capacity}).
\end{lemma}
\begin{IEEEproof}
 We use the code proposed by Han and Uchida \cite{Han00:cost}.
Then, from the property of the code, it holds that
\begin{eqnarray} \label{length}
c(\varphi_n^{\ast}({\bf x})) \leq - \frac{1}{\alpha_c} \log P_{X^n}({\bf x}) + \frac{\log 2}{\alpha_c} + c_{max},
\end{eqnarray}
for all $n=1,2,\cdots$, where $\varphi_n^{\ast}$ denotes the encoder of the code.
Furthermore, we set the decoder as the inverse mapping of $\varphi_n^{\ast}$.
Note that the code is a uniquely decodable code for {\it general} sources with countably infinite source alphabet .

Next, we shall evaluate the overflow probability of this code.
Set 
\[
A_n = \left\{{\bf x}\in {\cal X}^n \left|  z_n P_{X^n}(X^n) \leq K^{-\alpha_c \eta_n} \right. \right\},
\]
\begin{equation*} 
S_n = \left\{ {\bf x} \in {\cal X}^n \left|  c (\varphi_n^{\ast}({\bf x})) > \eta_n \right. \right\}.
\end{equation*}
Then, the overflow probability of this code is given by
\begin{eqnarray} \label{eq:1-1}
\varepsilon_n(\varphi^{\ast}_n, \eta_n) & = & \Pr \left\{ X^n \in S_n \right\} 
 \nonumber \\
& = & \sum_{{\bf x} \in S_n \cap A_n} P_{X^n}({\bf x}) + \sum_{{\bf x} \in S_n \cap A_n^c} P_{X^n}({\bf x}) \nonumber \\
& \leq & \Pr \left\{ X^n\!\in\!A_n \right\} + \sum_{{\bf x} \in S_n \cap A_n^c} P_{X^n}({\bf x}),
\end{eqnarray}
where $A^c$ denotes the complement set of the set $A$.

Since (\ref{length}) holds, for $\forall {\bf x}\in S_n$, we have
\begin{eqnarray*}
 - \frac{1}{\alpha_c} \log P_{X^n}({\bf x}) + \frac{\log 2}{\alpha_c} +c_{max}  > \eta_n.
\end{eqnarray*}
Thus, we have
\begin{eqnarray*}
 P_{X^n}({\bf x}) < K^{- \alpha_c \left(\eta_n - c_{max}\right) + \log 2},
\end{eqnarray*}
for $\forall {\bf x} \in S_n$.
Substituting the above inequality into (\ref{eq:1-1}), we have
\begin{align} \label{eq:1-2}
\lefteqn{\varepsilon_n(\varphi^{\ast}_n, \eta_n)} \nonumber \\
& < \Pr \left\{ X^n \in A_n \right\} + \sum_{{\bf x} \in S_n \cap A_n^c} K^{- \alpha_c \left(\eta_n - c_{max}\right)+ \log 2} \nonumber \\
& = \Pr \left\{ X^n \in A_n \right\} + \left| S_n \cap A_n^c \right|  K^{- \alpha_c \left(\eta_n - c_{max}\right)+ \log2}.
\end{align}
Here, from the definition of $A_n$, for $\forall {\bf x} \in A_n^c$, it holds that
\begin{equation*}
P_{X^n}({\bf x}) > \frac{K^{ - \alpha_c \eta_n}}{z_n}.
\end{equation*}
Thus, we have
\begin{align*}
1\!&\geq \sum_{{\bf x} \in A_n^c} P_{X^n}({\bf x}) > \sum_{{\bf x} \in A_n^c}  \frac{K^{ - \alpha_c \eta_n}}{z_n} = \left| A_n^c \right|  \frac{K^{ - \alpha_c \eta_n}}{z_n} .
\end{align*}
This mean that
\begin{equation} \label{eq:1-3}
\left| S_n \cap A_n^c \right| \leq  \left| A_n^c \right| < z_n K^{ \alpha_c \eta_n}.
\end{equation}
Substituting (\ref{eq:1-3}) into (\ref{eq:1-2}), we have
\begin{align*}
\varepsilon_n(\varphi^{\ast}_n, \eta_n)
 & \!<\! \Pr \left\{ X^n \!\in\! A_n \right\} \!+\! z_n K^{\alpha_c \eta_n} K^{- \alpha_c \left(\eta_n \!- c_{max}\right)+ \log 2} \nonumber \\
& \leq \Pr \left\{ X^n\in A_n \right\} + z_n K^{\alpha_c c_{max}+1},
\end{align*}
because $\log 2 \leq 1$ holds.
Therefore, we have proved the lemma. 
\end{IEEEproof}
%
%
%
%
%
\begin{lemma} \label{lemma2}
For any variable-length code and any sequence $\{\eta_n \}_{n=1}^\infty$, it holds that
\begin{eqnarray*}
\varepsilon_n(\varphi_n,\eta_n) \geq \Pr\left\{ P_{X^n}(X^n) \leq z_n K^{- \alpha_c \eta_n} \right\} - z_n,
\end{eqnarray*}
for $n=1,2,\cdots$ where $\{ z_n \}_{n=1}^\infty$ is a given sequence of an arbitrary number satisfying $z_i > 0$ for $i=1,2,\cdots$.
\end{lemma}
\begin{IEEEproof}
Let $\varphi_n$ be an encoder of variable-length code.
Set 
\[
B_n = \left\{ {\bf x} \in {\cal X}^n \left| P_{X^n}({\bf x}) \leq z_n K^{-\alpha_c \eta_n} \right. \right\},
\]
\begin{equation} \label{sn}
S_n = \left\{ {\bf x} \in {\cal X}^n \left|  c (\varphi_n({\bf x})) > \eta_n \right. \right\}.
\end{equation}

Then, 
we have
\begin{align}
\lefteqn{\Pr \left\{ P_{X^n}(X^n) \leq z_n K^{-\alpha_c \eta_n} \right\}} \nonumber \\
&= \sum_{{\bf x} \in B_n} P_{X^n}({\bf x})  = \sum_{{\bf x} \in B_n \cap S_n} P_{X^n}({\bf x}) + \sum_{{\bf x} \in B_n \cap S_n^c } P_{X^n}({\bf x}) \nonumber \\
& \leq \sum_{{\bf x} \in  S_n} P_{X^n}({\bf x}) + \sum_{{\bf x} \in B_n \cap S_n^c } P_{X^n}({\bf x}) \nonumber \\
& \leq \varepsilon_n(\varphi_n,\eta_n) + \sum_{{\bf x} \in B_n \cap S_n^c } P_{X^n}({\bf x}).
\end{align}
On the other hand, for $\forall {\bf x} \in B_n$ it holds that
\begin{equation*}
P_{X^n}({\bf x}) \leq z_n K^{-\alpha_c \eta_n}.
\end{equation*}
Thus, we have 
\begin{align} \label{eq:2-2}
\lefteqn{\Pr \left\{ P_{X^n}(X^n) \leq z_n K^{-\alpha_c \eta_n} \right\}} \nonumber \\
& \leq  \varepsilon_n(\varphi_n,\eta_n) + \sum_{{\bf x} \in B_n \cap S_n^c } z_n K^{-\alpha_c \eta_n} \nonumber \\
& \leq  \varepsilon_n(\varphi_n,\eta_n) + \left| B_n \cap S_n^c \right| z_n K^{-\alpha_c \eta_n}.
\end{align}
Here, from (\ref{kraft}), we have
\begin{align*}
1 & \geq  \sum_{{\bf x} \in {\cal X}^n}  K^{ - \alpha_c c(\varphi_x({\bf x})) }  \geq  \sum_{{\bf x} \in S_n^c}  K^{ - \alpha_c c(\varphi_x({\bf x})) }\\
& \geq  \sum_{{\bf x} \in S_n^c}  K^{ - \alpha_c \eta_n } = \left| S_n^c \right| K^{ - \alpha_c \eta_n} .
\end{align*}
This mean that
\begin{equation} \label{eq:2-2-1}
\left| B_n \cap S_n^c \right| \leq \left| S_n^c \right| \leq K^{ \alpha_c \eta_n}
\end{equation}

Hence, substituting (\ref{eq:2-2-1}) into (\ref{eq:2-2}), we have
\begin{align*}
\Pr \left\{ P_{X^n}(X^n) \!\leq\! z_n K^{ - \eta_n} \right\}
& \leq  \varepsilon_n(\varphi_n,\eta_n) \!+\!  K^{\alpha_c{\eta_n} } z_n K^{-\alpha_c \eta_n} \\
& =  \varepsilon_n(\varphi_n,\eta_n) +  z_n .
\end{align*}
Therefore, we have proved the lemma.
\end{IEEEproof}
%
%
%
%
\section{Infimum of $\varepsilon$-achievable threshold}
In this section, we determine $R(\varepsilon|{\bf X})$ for {\it general} sources.
Before showing the theorem, we define the function $F(R)$ as:
\[
F(R) \stackrel{\mathrm{def}}{=} \limsup_{n \to \infty} \Pr \left\{ \frac{1}{n\alpha_c} \log \frac{1}{P_{X^n}(X^n)} \geq R \right\}.
\]
The following theorem is one of our main results:
\begin{theorem} \label{theorem1} For $0 \leq \forall \varepsilon < 1$, it holds that 
\begin{eqnarray*}
R(\varepsilon|{\bf X}) = \inf \left\{ R \left| F(R) \leq \varepsilon \right. \right\}.
\end{eqnarray*}
\end{theorem}
\begin{IEEEproof}
The proof consists of two parts. \\
(Direct Part)
Let $R_0$ be as
\begin{equation*}
R_0 = \inf \left\{ R \left| F(R) \leq \varepsilon \right. \right\},
\end{equation*}
for short.
Then, in this part we prove that
\begin{equation*}
R(\varepsilon|{\bf X}) \leq R_0 + \gamma,
\end{equation*}
for any $\gamma >0$ by showing that $R_0$ is an $\varepsilon$-achievable overflow threshold for the source.
Let $\eta_n$ be as $\eta_n = n(R_0 + \gamma)$, then from Lemma \ref{lemma1} there exists a variable-length encoder $\varphi_n$ that satisfies
\begin{multline*}
\varepsilon_n(\varphi_n,nR_0) \\
 \!<\! \Pr\left\{ z_n P_{X^n}(X^n)\! \leq\! K^{\!-n\alpha_c (R_0+\gamma)} \right\} + z_n K^{\alpha_c c_{max} + 1},
\end{multline*}
for $n=1,2,\cdots$. 
Thus, we have
\begin{align*}
\varepsilon_n(\varphi_n,nR_0)   <  & \Pr\left\{ \frac{1}{n \alpha_c} \log \frac{1}{P_{X^n}(X^n)} \!\geq\! R_0 \!+\! \gamma \!+\! \frac{\log z_n}{n\alpha_c}  \right\}  \\
& + z_n K^{\alpha_c c_{max}+1},
\end{align*}
for $n=1,2,\cdots$.
Notice that $z_n > 0$ is an arbitrary number. Set $z_n = K^{-\sqrt{n} \gamma}$, 
then we have
\begin{align*}
\varepsilon_n(\varphi_n,nR_0) & < \Pr\left\{ \frac{1}{n \alpha_c} \log \frac{1}{ P_{X^n}(X^n)}\geq R_0 \!+\! \gamma \!-\! \frac{\sqrt{n} \gamma}{n\alpha_c } \right\}  \\
 & \ \ \ +K^{-\sqrt{n}\gamma + \alpha_c c_{max}+1} \\
& < \Pr\left\{ \frac{1}{n \alpha_c} \log \frac{1}{ P_{X^n}(X^n)}\geq R_0 \right\}  \\
 & \ \ \ +K^{-\sqrt{n}\gamma + \alpha_c c_{max}+1},\end{align*}
for sufficiently large $n$, because $\gamma> \frac{ \gamma}{\sqrt{n} \alpha_c}$ as $n \to \infty$.
Thus, since $\alpha_c$ and $c_{max}$ are positive constants, by taking $\limsup_{n \to \infty}$, we have
\begin{align*}
\lefteqn{\limsup_{n \to \infty} \varepsilon_n(\varphi_n,nR_0)} \\ 
& \leq\!  \limsup_{n \to \infty} \Pr\left\{ \frac{1}{n \alpha_c} \log \frac{1}{ P_{X^n}(X^n)} \!\geq\! R_0  \right\}.
\end{align*}

Hence, from the definition of $R_0$ we have
\begin{eqnarray*}
\limsup_{n \to \infty} \varepsilon_n(\varphi_n,nR_0) & \leq & \varepsilon.
\end{eqnarray*}
Therefore, the direct part has been proved. \\
%
%
(Converse Part) \\
Assuming that $R_1$ satisfying
\begin{equation} \label{eq:4-2-0}
R_1 < \inf \left\{ R \left| F(R) \leq \varepsilon \right. \right\},
\end{equation}
is an $\varepsilon$-achievable overflow threshold. Then we shall show a contradiction.

Let $\eta_n$ be as $\eta_n = nR_1$. Then, from Lemma \ref{lemma2} for any sequence $\{ z_n \}_{n=1}^\infty$ ( $z_i > 0$ $i=1,2,\cdots$) and any variable-length encoder we have
\begin{eqnarray*}
\varepsilon_n(\varphi_n, nR_1) > \Pr\left\{ P_{X^n}(X^n) \leq z_n K^{ - n\alpha_cR_1} \right\} - z_n,
\end{eqnarray*}
for $n=1,2,\cdots$.
Thus, for any variable-length encoder it holds that for each $n=1,2,\cdots$
\begin{align*}
\varepsilon_n(\varphi_n,nR_1) 
&\!>\! \Pr\left\{ \frac{1}{n\alpha_c} \log \frac{1}{P_{X^n}(X^n)}\!\geq\!R_1\!-\! \frac{\log z_n}{n} \right\}\!-\!z_n.
\end{align*}
Set $z_n = K^{-n \gamma}$, where $\gamma >0$ is a small constant that satisfies
\begin{equation} \label{eq:4-2-1}
R_1 + \gamma < \inf \left\{ R \left| F(R) \leq \varepsilon \right. \right\}.
\end{equation}
Since we assume that (\ref{eq:4-2-0}) holds, it is obvious that there exists $\gamma >0$ that satisfies the above inequality.
Then, we have
\begin{align*}
\varepsilon_n(\varphi_n,nR_1) > & \Pr\left\{ \frac{1}{n\alpha_c} \log \frac{1}{P_{X^n}(X^n)} \geq R_1\!+\!\gamma \right\}  - K ^{ -n\gamma},
\end{align*}
for any variable-length encoder.
Hence, we have
\begin{align*}
\lefteqn{\limsup_{n \to \infty}\varepsilon_n(\varphi_n,nR_1)} \\ 
& \geq  \limsup_{n \to \infty} \Pr\left\{ \frac{1}{n} \log \frac{1}{P_{X^n}(X^n)} \geq R_1 + \gamma \right\} 
> \varepsilon,
\end{align*}
where the last inequality is derived from (\ref{eq:4-2-1}) and the definition of $F(R)$.
This is a contradiction and the converse part has been proved. 
\end{IEEEproof}
\begin{remark} \label{remark1}
In the analysis of $\varepsilon$-fixed-length source coding for {\it general} sources, the function 
\[
F(R) = \limsup_{n \to \infty} \Pr \left\{ \frac{1}{n\alpha_c} \log \frac{1}{P_{X^n}(X^n)} \geq R \right\}.
\]
is used in the case that $\alpha_c =1$ \cite{Steinberg1996} (see, also \cite[Theorem 1.6.1]{Han}).
This suggests a deep relationship between the overflow probability of codeword {\it length} in variable-length coding and the error probability in fixed-length coding.
The similar relationship in the {\it second-order} case is also clarified by Theorem \ref{theorem2} in the following section and \cite[Theorem 3]{Hayashi}.
\end{remark}

From the above theorem, we can show a corollary. Before describing the corollary, we define the spectral sup-entropy rate \cite{Han}\footnote{%
For any sequence $\{Z_n \}_{n=1}^{\infty}$ of real-valued random variables, we define the limit superior in probability of $\{Z_n \}_{n=1}^{\infty}$ by
$\mbox{p-}\limsup_{n \to \infty} Z_n = \inf \left\{ \beta | \lim_{n \to \infty} \Pr \{Z_n > \beta \} = 0 \right\}$ (cf.\cite{Han}) .
}
:
\begin{equation*}
\overline{H}({\bf X}) \stackrel{\mathrm{def}}{=} \mbox{p-}\limsup_{n \to \infty} \frac{1}{n} \log \frac{1}{P_{X^n}(X^n)}.
\end{equation*}
Then, the following corollary holds.
\begin{corollary} \label{coro1}
\begin{equation} \label{supentropy}
R(0|{\bf X}) = \frac{1}{\alpha_c}\overline{H}({\bf X}).
\end{equation} 
\end{corollary}
\section{Infimum of $(\varepsilon,a)$-achievable threshold}
So far, we have considered the {\it first-order} achievable threshold. 
In this section, we consider the {\it second-order} achievability.
In the {\it second-order} case, the infimum $(\varepsilon,a)$-achievable overflow threshold for general sources is also determined by using Lemma \ref{lemma1} and Lemma \ref{lemma2}.

We define the function $F_a(R)$ given $a$ as follows, which is correspondence with the function $F(R)$ in the {\it first-order} case.
\begin{equation*}
F_a(L) \stackrel{\mathrm{def}}{=} \limsup_{n \to \infty} \Pr \left\{ \frac{ - \log P_{X^n}(X^n) -n \alpha_c a }{\sqrt{n}\alpha_c} \geq  {L} \right\}.
\end{equation*}
Then, we have 
\begin{theorem} \label{theorem2} For $0 \leq \forall \varepsilon < 1$, it holds that 
\begin{eqnarray*}
L(\varepsilon,a|{\bf X}) = \inf \left\{ L \left| F_a(L) \leq \varepsilon \right. \right\}.
\end{eqnarray*}
\end{theorem}
\begin{IEEEproof}
The proof is similar to the proof of Theorem \ref{theorem1}. Hence, we omit the proof.
\end{IEEEproof}

The above theorem is valid for {\it general } sources.
Hence Theorem \ref{theorem2} is a quite general result.
However, in general to compute the function $L\left( \varepsilon,a|{\bf X}\right)$ is hard.
Next, we consider a simple case such as an i.i.d. source and we address the above quantity specifically.

For an i.i.d. source, from Corollary \ref{coro1}, we are interested in $L\left( \left. \varepsilon,\frac{1}{\alpha_c} H(X)\right|{\bf X}\right) $.
To specify this quantity  for an i.i.d. source, we need to introduce the variance of self-information as follows:
\begin{equation*}
\sigma^2(X) \stackrel{\mathrm{def}}{=} E \left(-\log P_{X}(X) - H(X)\right)^2.
\end{equation*}
In this paper, we assume that the above variance exists.
Then, from Theorem \ref{theorem2} we obtain the following theorem.
\begin{theorem}
For any i.i.d. source, it holds that
\begin{eqnarray*} 
L\left( \left. \varepsilon,\frac{1}{\alpha_c} H(X) \right|{\bf X} \right) = \frac{1}{{\alpha_c}}\sqrt{\sigma^2(X)} \Phi^{-1}(1 - \varepsilon),
\end{eqnarray*}
where $\Phi^{-1}$ denotes a inverse function of $\Phi$ and 
$\Phi(T)$ is the Gaussian cumulative distribution function with mean $0$ and variance $1$, that is, $\Phi(T)$ is given by
\begin{eqnarray} \label{eq:5-1}
\Phi(T) & = & \int_{-\infty}^{T} \frac{1}{\sqrt{2\pi }}\exp\left[ -\frac{y^2}{2} \right] dy.
\end{eqnarray}
\end{theorem}
\begin{IEEEproof}
From the definition of $F_a(L)$,  we have
\begin{align*}
\lefteqn{F_{H(X)/\alpha_c}(L)} \\
& =  \limsup_{n \to \infty} \Pr \left\{ \frac{ - \log P_{X^n}(X^n) -n H(X) }{\sqrt{n}\alpha_c } \geq  {L} \right\} \\
& =  \limsup_{n \to \infty} \Pr \left\{ \frac{ - \log P_{X^n}(X^n) - H(X) }{\sqrt{n\sigma^2(X)}}\geq  \frac{L \alpha_c }{\sqrt{\sigma^2(X)}} \right\} .
\end{align*}
On the other hand, since we consider the i.i.d. source
, from the asymptotic normality (due to the central limit theorem) it holds that
\begin{multline*}
 \lim_{n \to \infty} \Pr\left\{ \frac{\!-\! \log P_{X^n}(X^n)\!-\!nH(X) }{\sqrt{n \sigma^2(X)}}\leq U \right\}\\
  =  \int_{-\infty}^{U} \frac{1}{\sqrt{2\pi }}\exp\left[ -\frac{y^2}{2} \right] dy.
\end{multline*}
Thus, we have
\begin{align*}
F_{H(X)/\alpha_c}(L) = \int^{\infty}_{\frac{L \alpha_c}{\sqrt{\sigma^2(X)}}} \frac{1}{\sqrt{2\pi }}\exp\left[ -\frac{y^2}{2} \right] dy.
\end{align*}
Thus, $L\left( \left. \varepsilon,\frac{1}{\alpha_c}H(X) \right| {\bf X} \right) $ is given by
\begin{align*} 
\lefteqn{L\left( \left. \varepsilon,\frac{1}{\alpha_c}H(X) \right|{\bf X} \right)} \\
& =  \inf \left\{ L \left| \int^{\infty}_{\frac{L\alpha_c}{\sqrt{\sigma^2(X)}}} \frac{1}{\sqrt{2\pi }}\exp\left[ -\frac{y^2}{2} \right] dy  \leq \varepsilon \right. \right\} \\
& = \inf \left\{ L \left| 1- \Phi \left( \frac{L\alpha_c}{\sqrt{\sigma^2(X)}} \right) \leq \varepsilon \right. \right\}
\end{align*}
Since $\Phi \left( \frac{L\alpha_c}{\sqrt{\sigma^2(X)}} \right)$ is a continuous function and monotonically increases as $L$ increases, we have
\[
\frac{L\left( \left. \varepsilon,\frac{1}{\alpha_c}H(X) \right|{\bf X} \right) \alpha_c}{\sqrt{\sigma^2(X)}} = \Phi^{-1}(1-\varepsilon).
\]
Therefore, the proof has been completed. 
\end{IEEEproof}
%
%
%
%
%
%
%
%
%
%
%
%
%
%
\section{Conclusion}
We have so far dealt with the overflow probability of variable-length coding with cost function for {\it general} sources.
The overflow probability is important not only from the theoretical viewpoint but also from the engineering point of view.
However, there is few research on the overflow probability of variable-length coding for {\it general} sources, even though the equal cost function is assumed. 
Hence, our attempt to analyze the overflow probability  
is meaningful.
%

We have revealed that, as shown in the proofs of the present paper, the {\it information-spectrum} approach is substantial in the analysis of the overflow probability of variable-length coding with cost function. 
An application to more general cost function is a future work.

Finally, as described in Remark \ref{remark1} there is a deep relationship between the overflow probability of codeword {\it length} in variable-length coding and the error probability in fixed-length coding. It is an interesting problem to consider the relation of the overflow probability of codeword {\it cost} to the fixed-length coding. 
\bibliographystyle{IEEEtran}
%

\end{document}